\documentclass[fleqn,12pt,twoside]{article}
\usepackage{espcrc1}

\usepackage{graphicx}

\title{Physical Baryon Resonance Spectroscopy from Lattice QCD}
\author{D. Morel, B. Crouch, D. B. Leinweber, and A. W. Thomas \\
\addressmark{Centre for the Subatomic Structure of Matter \& Department of 
Physics,\\ University of Adelaide, SA,\ 5005, Australia }}
       
\begin{document}

\def\half{{\textstyle{1\over2}}}
\def\thalf{{\textstyle{3\over2}}}

\maketitle

\vspace{-7cm}
\null \hfill ADP-03-131/T567
\vspace{7cm}
\vspace{-24pt}

\begin{abstract}
We complement recent advances in the calculation of the masses of
excited baryons in quenched lattice QCD with finite-range regulated
chiral effective field theory enabling contact with the physical quark
mass region. We examine the $P$-wave contributions to the low-lying
nucleon and delta resonances.
\end{abstract}
\vspace{-1mm}
\section{INTRODUCTION}
\vspace{-1mm}
There has been enormous progress in the calculation of the masses of
excited baryons in quenched lattice QCD
~\cite{Allton:wc}-\cite{Zanotti:2002nk} as a consequence of improved
computing resources and the development of spin and parity projection
methods.  At present, established fitting techniques have successfully
extracted resonance masses for quark masses or (equivalently, through
the Goldberger-Treiman relation) pion masses in the range $m_\pi^2 >
0.3$ GeV$^2$. The difficulty lies in maintaining a clear signal for
the effective mass of the baryon resonance as the quarks become
light. In this region the masses of the first negative and positive
parity excited states obtained from standard interpolating fields
follow the expectations of the naive harmonic oscillator based quark
model with roughly equal spacing between the ground state nucleon, the
first $1/2^-$ and the first $1/2^+$ excited states.

Quite apart from the ordering of excited states, the chiral
extrapolation of hadron masses is a topic of current interest because
of the well known non-analytic behaviour with quark mass and the need
to incorporate the consequent, model independent constraints of chiral
symmetry. For the octet and decuplet baryons as well as the
$\rho$-meson this problem has been studied extensively and the
technique for extracting physical masses with small systematic
uncertainties is well
understood~\cite{Leinweber:1999ig}-\cite{Cloet:2003jm}. Our aim is to
generalise those results to the baryon excited states currently
accessible on the lattice.
\vspace{-1mm}
\section{CHIRAL PERTURBATION THEORY AND CHIRAL EXPANSION}
\vspace{-1mm}
In the context of the extrapolation of lattice data, chiral
perturbation theory provides a functional form applicable as the quark
mass vanishes or, equivalently $m_\pi^2\rightarrow 0$. Goldstone boson
loops play an important role in the theory as they give rise to
non-analytic behavior as a function of quark mass. All hadron
properties receive contributions involving these loops. For $m_\pi<$
400-500 MeV, Goldstone loops lead to rapid, non-analytic variation
with $m_q$. This non-analyticity must therefore be incorporated in any
extrapolation scheme.

Chiral effective field theory gives a formal expansion of a baryon
mass about the chiral limit
\vspace{-2mm}
\begin{equation}
M_B = a_0 + a_2 m_\pi^2 + a_4 m_\pi^4 + \sum_{B'}
\sigma_{BB'\pi}(m_\pi,\Lambda) + \ldots
\end{equation}
\vspace{-3mm}
where $a_i$ are the coefficients of the analytic terms, $\Lambda$ is a
parameter associated with the regularisation, and $\sigma_{BB'\pi}$
are pion-induced self-energy contributions. Here we consider $P$-wave
pion-baryon contributions, focusing on the most important baryon
states as determined by the strength of the meson-baryon coupling
constants (see Ref.~\cite{Morel:2002vk} for details).

\begin{figure}[htb]
\vspace{-4.5mm}
\begin{minipage}[t]{80mm}
\includegraphics[width=79mm]{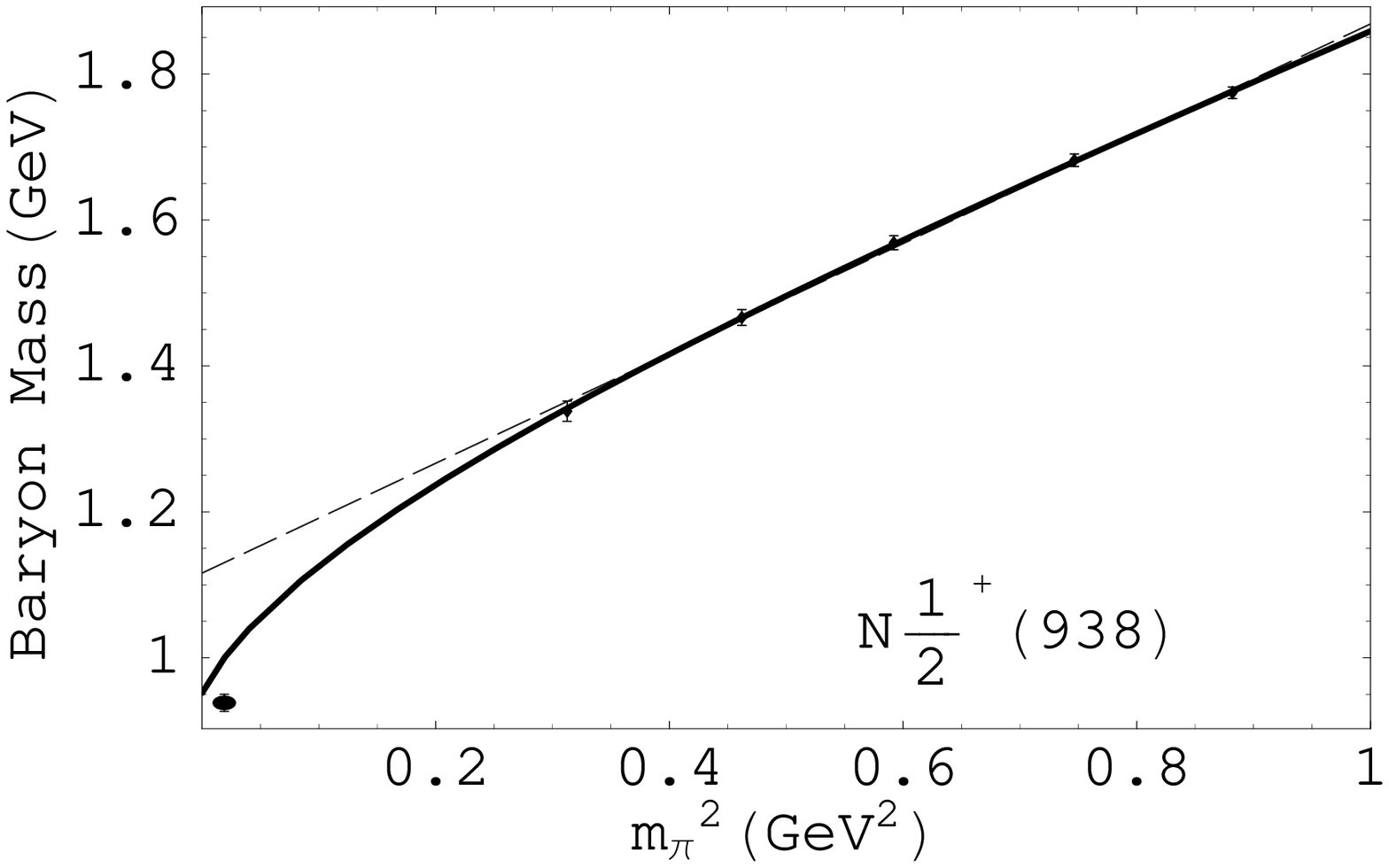}
\end{minipage}
\vspace{3mm}
\begin{minipage}[t]{80mm}
\includegraphics[width=79mm]{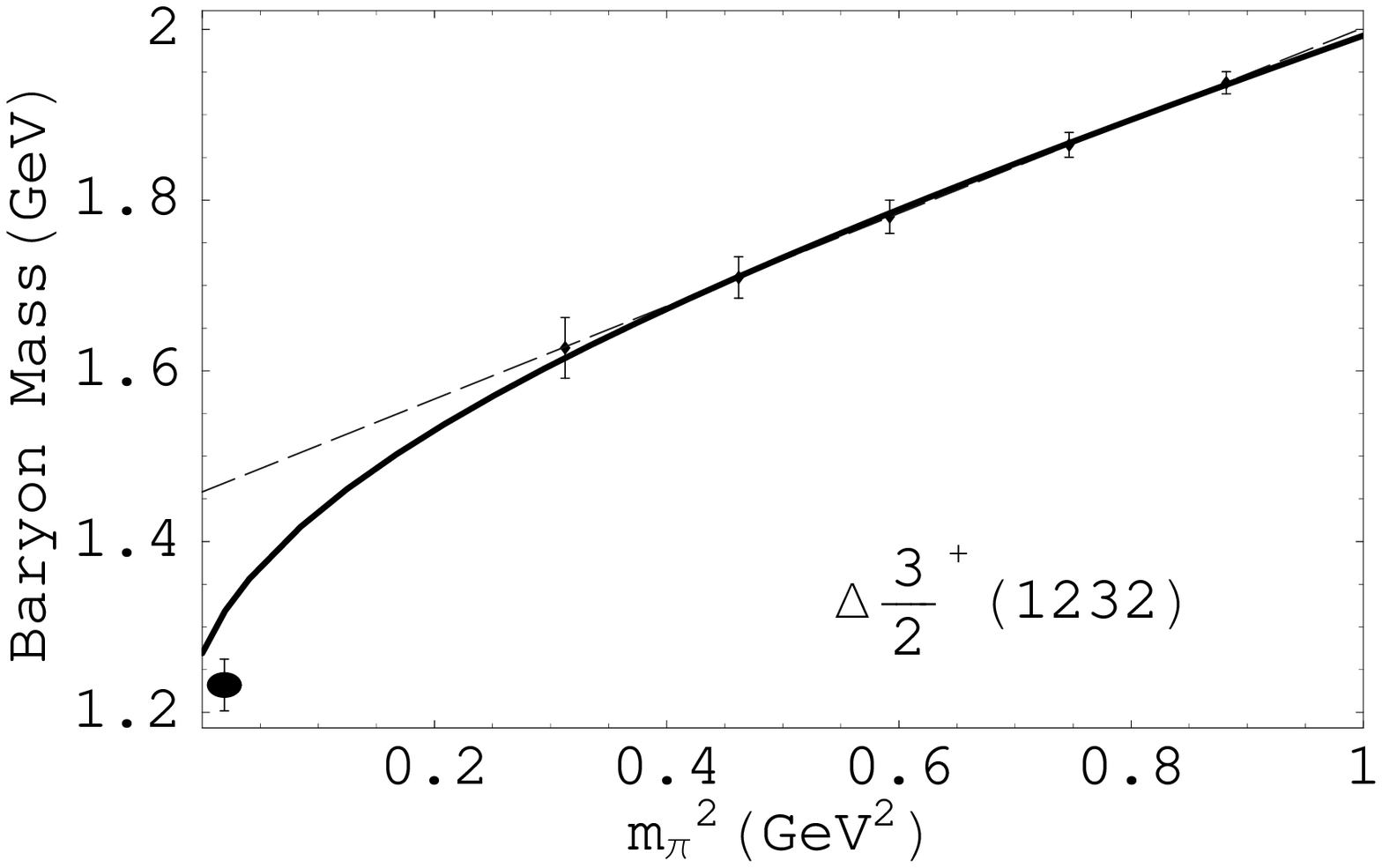}
\end{minipage}
\begin{minipage}[t]{80mm}
\includegraphics[width=79mm]{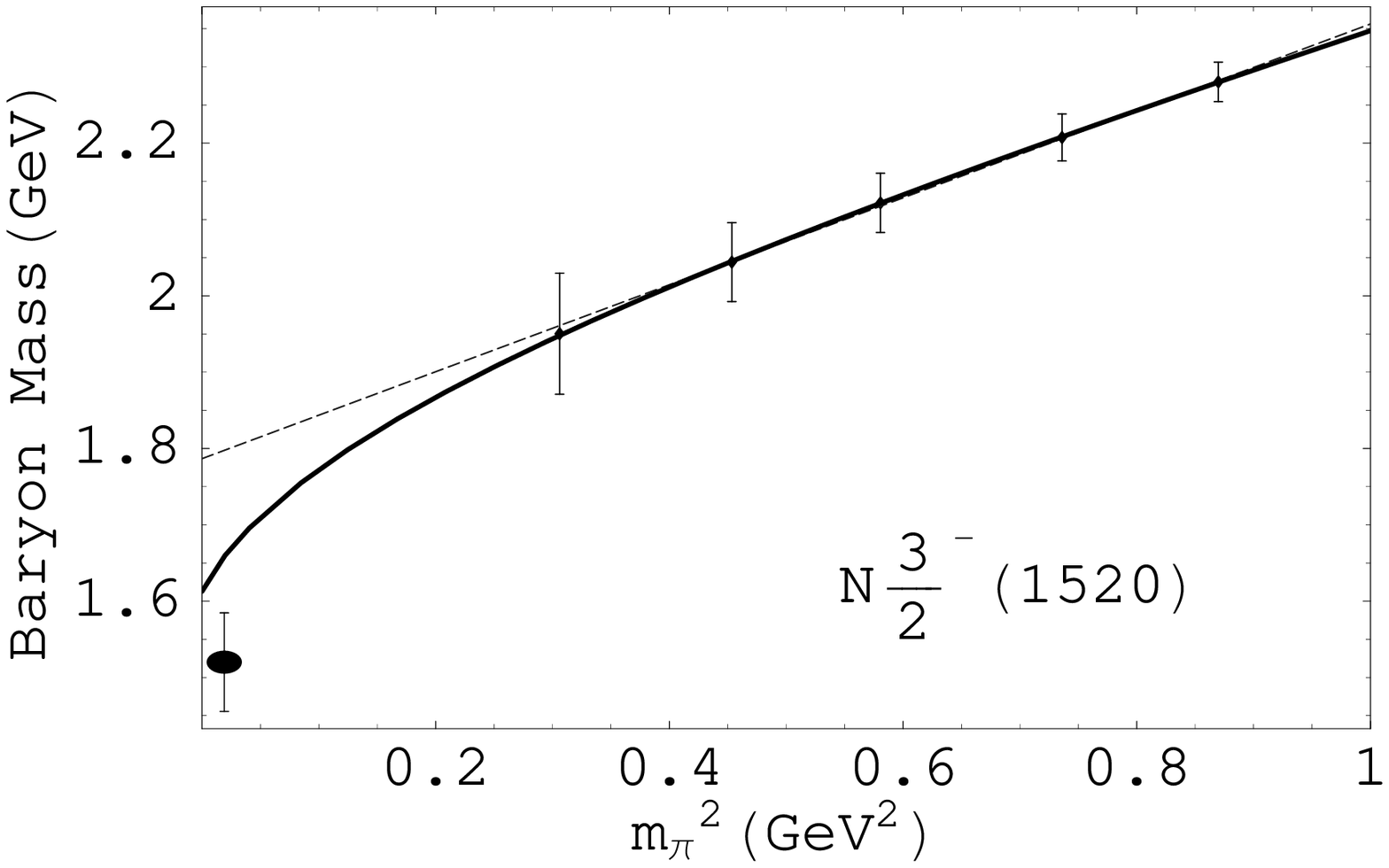}
\end{minipage}
\vspace{3mm}
\begin{minipage}[t]{80mm}
\includegraphics[width=79mm]{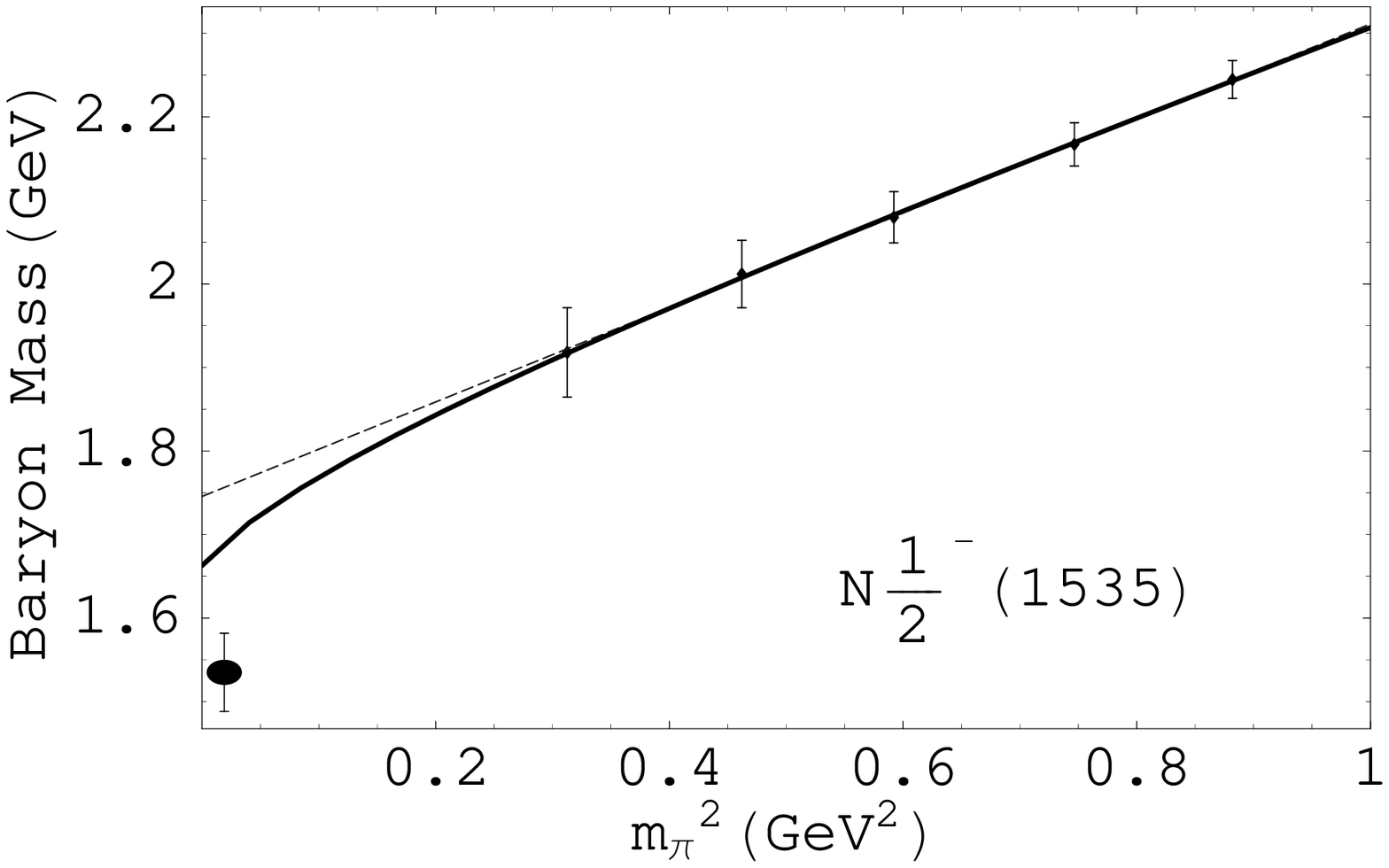}
\end{minipage}
\begin{minipage}[b]{80mm}
\includegraphics[width=79mm]{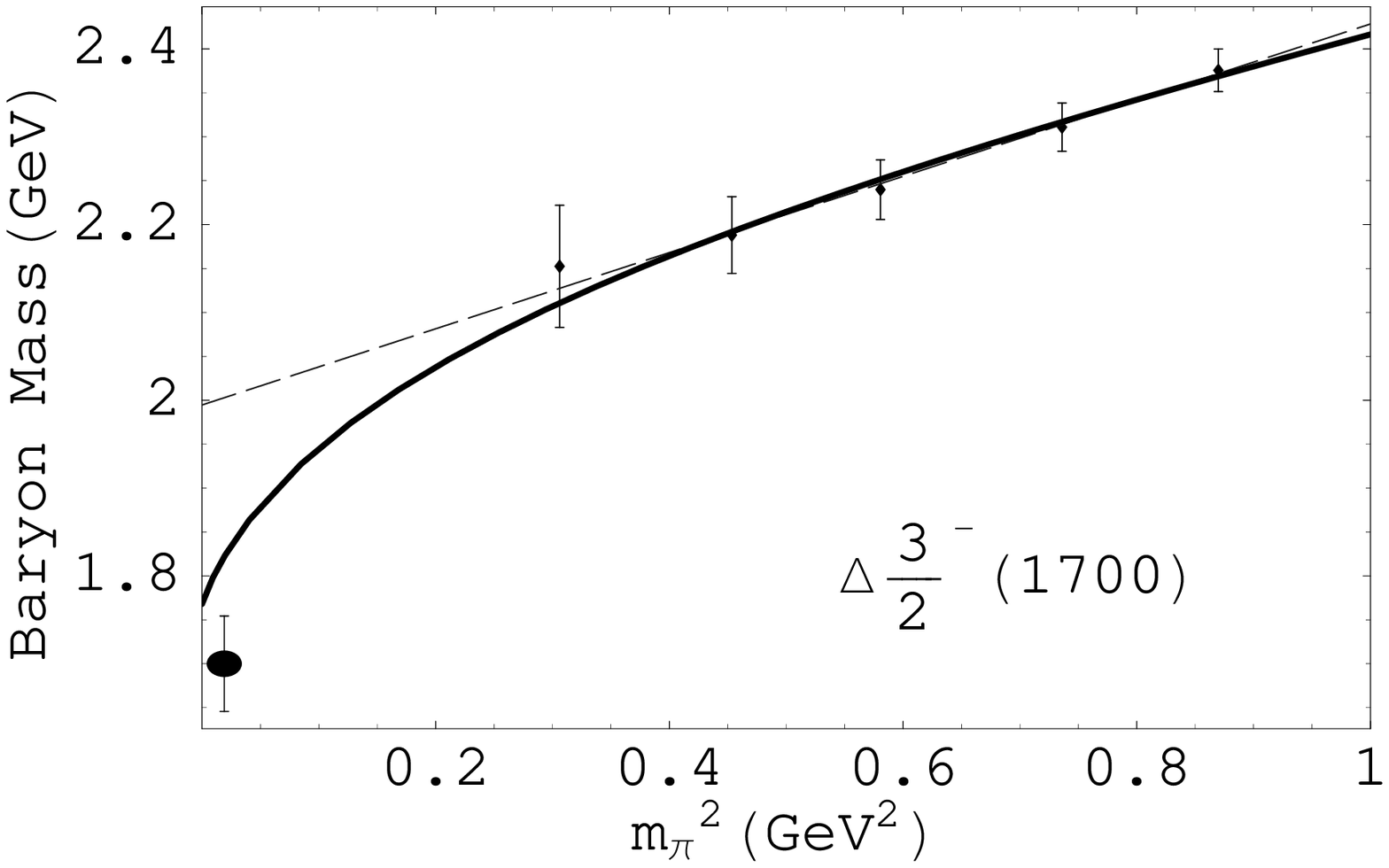}
\end{minipage}
\begin{minipage}[b]{80mm}
\includegraphics[width=79mm]{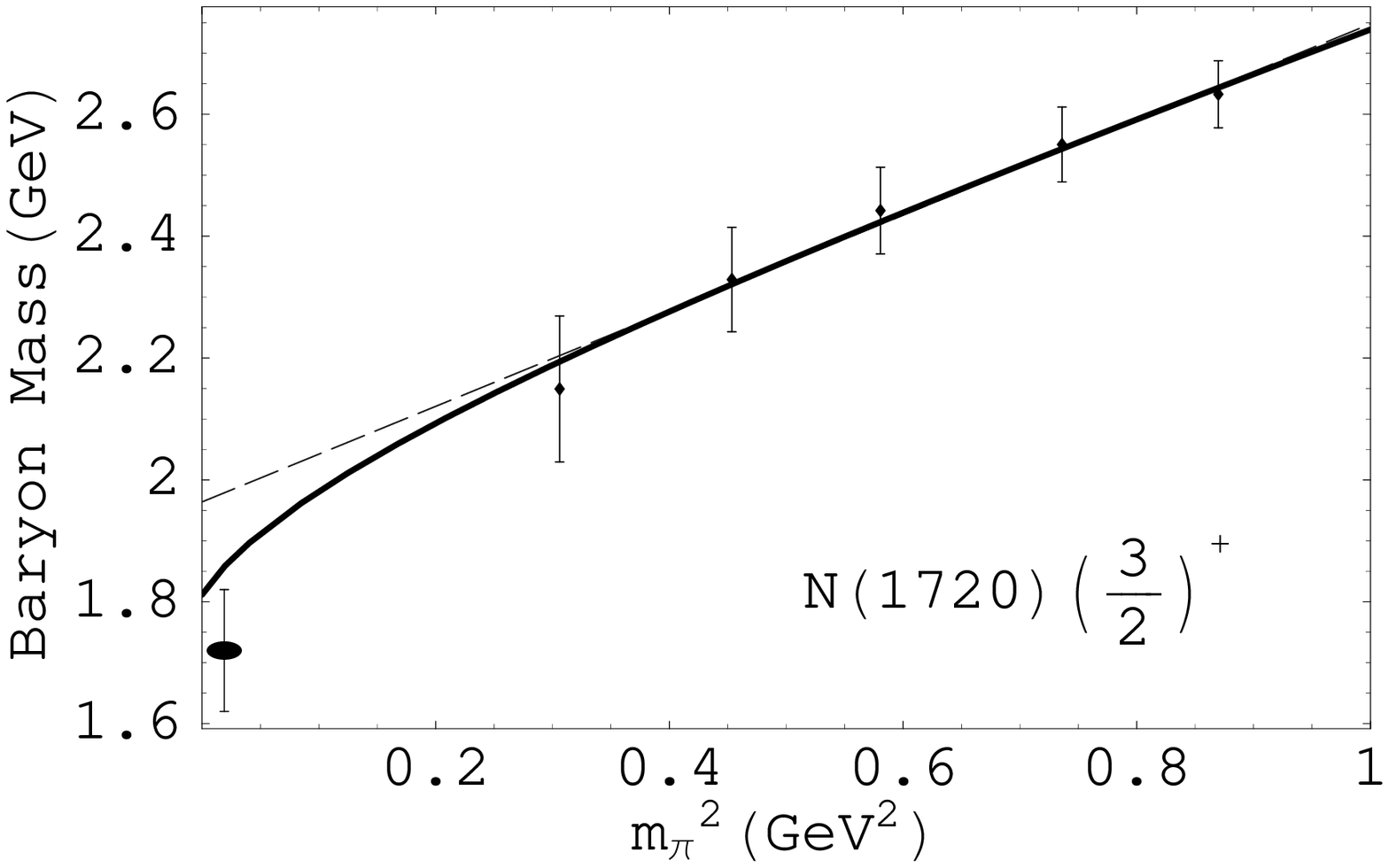}
\end{minipage}
\vspace{-14mm}
\caption{\label{fig} Extrapolation of lattice data for selected baryon ground states and resonances. Symbols are explained in the text.}
\end{figure}

Figure~\ref{fig} shows the extrapolation of CSSM Lattice collaboration
\cite{Melnitchouk:2002eg,Zanotti:2002nk,Zanotti:2003fx} data
for ground-state and low-lying excited baryon states. Dashed lines
represent naive linear extrapolations while solid lines incorporate
the non-analytic variation with $m_q$. Experimental
masses are shown as filled ellipses.

\section{CONCLUSION \& OUTLOOK}

We have presented the first quantitative analysis of the baryon
resonance spectrum from lattice QCD in which the chiral non-analytic
behavior of QCD is incorporated. It is interesting to note that a
small downward shift in the lattice results upon unquenching (as
suggested in Ref.~\cite{Zanotti:2003fx}) would provide remarkable
agreement with experiment. The incorporation of chiral non-analytic
behavior inverts the $N\half^-(1535)$ and $N\thalf^- (1520)$ ordering
when compared to the naive linear extrapolation, in accord with
experiment.

Encouraged by these early successes, work is now in progress to
include additional self-energy contributions. $S$-wave contributions
are of particular interest due to the non-analytic structure of their
contributions. Ultimately, quenched chiral perturbation theory will be
formulated providing the opportunity to quantitatively estimate the
predictions of full QCD.


\begin{thebibliography}{99}

\bibitem{Allton:wc}
C.~R.~Allton {\it et al.}  
[UKQCD Collaboration],
Phys.\ Rev.\ D {\bf 47}, 5128 (1993).
%
\bibitem{Leinweber:1994nm}
D.~B.~Leinweber,
Phys.\ Rev.\ D {\bf 51}, 6383 (1995).
%
\bibitem{Lee:1998cx}
F.~X.~Lee and D.~B.~Leinweber,
Nucl.\ Phys.\ Proc.\ Suppl.\ {\bf 73}, 258 (1999).
%
\bibitem{Richards:2001bx}
D.~G.~Richards,
M.~Gockeler, R.~Horsley, D.~Pleiter, P.~E.~Rakow,
G.~Schierholz and C.~M.~Maynard
Nucl.\ Phys.\ Proc.\ Suppl.\ {\bf 109}, 89 (2002).
%
\bibitem{Gockeler:2001db}
M.~Gockeler, R.~Horsley, D.~Pleiter, P.~E.~Rakow, G.~Schierholz,
C.~M.~Maynard and D.~G.~Richards,
Phys.\ Lett.\ B {\bf 532}, 63 (2002).
%
\bibitem{Dong:2003zf}
S.~J.~Dong, T.~Draper, I.~Horvath, F.~X.~Lee, K.~F.~Liu, N.~Mathur and J.~B.~Zhang,
hep-ph/0306199.
%
\bibitem{Sasaki:2001nf}
S.~Sasaki, T.~Blum and S.~Ohta,
Phys.\ Rev.\ D {\bf 65}, 074503 (2002).

\bibitem{Melnitchouk:2002eg}
W.~Melnitchouk {\it et al.}, 
Phys.\ Rev.\ D {\bf 67}, 114506 (2003),
hep-lat/0210042.
%
\bibitem{Zanotti:2002nk}
J.~M.~Zanotti, {\it et al.},
hep-lat/0210043.

\bibitem{Leinweber:1999ig}
D.~B.~Leinweber, A.~W.~Thomas, K.~Tsushima and S.~V.~Wright,
Phys.\ Rev.\ D {\bf 61}, 074502 (2000),
%
Phys.\ Rev.\ D {\bf 64}, 094502 (2001).
%
\bibitem{Young:2002cj}
R.~D.~Young, D.~B.~Leinweber, A.~W.~Thomas and S.~V.~Wright,
Phys.\ Rev.\ D {\bf 66}, 094507 (2002).
%
\bibitem{Young:2002ib}
R.~D.~Young, D.~B.~Leinweber and A.~W.~Thomas,
Prog.\ Part.\ Nucl.\ Phys. {\bf 50} (2003) 399.
%
\bibitem{Young:new}
D.~B.~Leinweber, A.~W.~Thomas and R.~D.~Young,
hep-lat/0302020. 
%
\bibitem{Cloet:2003jm}
I.~C.~Cloet, D.~B.~Leinweber and A.~W.~Thomas,
Phys.\ Lett.\ B {\bf 563} (2003) 157.

\bibitem{Morel:2002vk}
D.~Morel and S.~Capstick,
nucl-th/0204014;
%
S.~Capstick and N.~Isgur,
Phys.\ Rev.\ D {\bf 34}, 2809 (1986);
%
S.~Capstick and W.~Roberts,
Phys.\ Rev.\ D {\bf 47}, 1994 (1993),
%
Phys.\ Rev.\ D {\bf 49}, 4570 (1994).

\bibitem{Zanotti:2003fx}
J.~M.~Zanotti, D.~B.~Leinweber, A.~G.~Williams, J.~B.~Zhang,
W.~Melnitchouk and S.~Choe,
hep-lat/0304001.

\end{thebibliography}
\end{document}